\documentclass[aps,prl,reprint,notitlepage,groupedaddress]{revtex4-2}
\usepackage{graphicx,psfrag,color}
\usepackage{bm}
\usepackage{natbib}
\usepackage{amssymb,amsmath,esint}
\usepackage{pifont}
\usepackage[mathscr]{eucal}
\usepackage{hyperref}
\hypersetup{breaklinks=true,colorlinks = true, linkcolor = blue, citecolor = blue, urlcolor = blue}

\begin{document}


\title{
Clogging-flowing transition of granular media in a two-dimensional vertical pipe
}
\author{Y. Zhou$^{1}$}
\email{yixian.zhou@ncepu.edu.cn}
\author{M. Li$^1$}
\author{Y. Guan$^1$}
\author{Y. Wang$^1$}
\author{Y. Liu$^1$}
\author{Z. Zou$^{2}$}
\email{zou\_zhenhai@sina.com}
\affiliation{1: Beijing Key Laboratory of Passive Safety Technology for Nuclear Energy, North China Electric Power University, Beijing 102206, China}

\affiliation{2: State Key Laboratory of Marine Thermal Energy and Power, Wuhan, Hubei 430205, China}

\begin{abstract}
We experimentally and numerically investigate the clogging behavior of granular materials in a two-dimensional vertical pipe. The nonmonotonicity of clogging probability found in a cylindrical vertical pipe~[L\'opez et al., Phys. Rev. E 102, 010902 (2020)] is also observed in the two-dimensional case. We numerically demonstrate that the clogging probability strongly correlates with the friction coefficient $\mu$ in addition to the pipe-to-particle diameter ratio $D/d$. We thus construct a clogging-flowing $D/d$-$\mu$ phase diagram within the $2<D/d<3$ range. Finally, by analyzing the geometrical arrangements of particles and using a simple analysis of forces and torques, we are able to predict the clogging-flowing transition in the $D/d$-$\mu$ phase diagram and explain the mechanism of the observed counterintuitive nonmonotonicity in more detail. We demonstrate that for pipe clogging, friction-mediated force and torque equilibrium are essential for arch formation.

\end{abstract}

\maketitle


\textit{Introduction.} Clogging behavior can be frequently observed in various systems, such as sheep herds~\cite{ZUR14,GAR2015}, pedestrian crowds \cite{HER2024,ECH2022}, colloidal suspensions~\cite{DRE2017,HID2018}, microswimmers \cite{AIA2022}, and granular media \cite{TO01,Zuriguel2005, POU07,Endo2017,Yu2021}. 
Compared to the extensive research on converging geometries like hoppers and silos, there is relatively little research work on clogging in vertical pipes with a uniform cross section. This phenomenon poses severe risks in various industry sectors \cite{Verbucheln2015}, such as food and pharmaceutical processing, pneumatic conveying \cite{Altino2025}, underground mining \cite{Hadjigeorgiou2005, Janda2015}, and nuclear power systems, particularly during the transport of spherical fuel elements in high-temperature gas-cooled reactors \cite{Zhang2009}.

In hopper flows, many parameters related to the clogging phenomenon have been investigated, for instance, the outlet size \cite{TO01, Zuriguel2003, Janda2008,Garcimartin2010}, number of particles \cite{Yu2021,Janda2008}, friction coefficient \cite{POU07}, particle stiffness \cite{Xia2017,Tao2021,Alborzi2022}, hopper angle \cite{Yu2021,Lopez2019}, driving force \cite{Arevalo2016,Arevalo2014}, width of silo \cite{Gella2017}, and shape of particle \cite{Zuriguel2005, Ashour2017}. Among these parameters, it is widely recognized that the aspect ratio of the outlet size $D$ to the particle size $d$ is a crucial determinant of clogging probability $J$ \cite{TO01, Zuriguel2005}. In a seminal study, To et al. \cite{TO01} demonstrated that $J(D/d)$ can be quantitatively obtained by using the restricted random walker method (RRWM). This approach captures the essential geometric constraints of arch formation that prevent particle overlap, without calculating interparticle forces.

Several experimental and numerical studies in vertical pipes have shown that clogging can happen when the diameter of the pipe $D$ is not much larger than the particle diameter $d$ \cite{Lopez2020,Janda2015,Hadjigeorgiou2007NumericalIO,Verbucheln2015,Verbucheln2017,KVAPIL1965}. Recently L\'opez et al. \cite{Lopez2020} report experimental evidence of clogging of spherical particles in a cylindrical vertical pipe. They found that, unlike in silos, the clogging probability $J$ in vertical narrow tubes does not decrease monotonically with the pipe-to-particle diameter ratio $D/d$. By using discrete-element method (DEM) simulations, they revealed that the absence of clogging at specific $D/d$ (e.g., $D/d = 2, 2.43, 3$) results from the spontaneous development of an ordered structure (crystallization) within the packing. However, a predictive framework to determine the $D/d$ values for the transition to crystallization has not yet been established. Notably, near $D/d = 2.21$, the experimental results showed a low clogging probability, while simulations predicted a high probability accompanied by a disordered structure. This discrepancy may be attributed to differences in the friction coefficients between the experiments and the simulations. Therefore, the underlying physical mechanisms of clogging in pipes warrant further investigation, particularly the role of friction coefficients and the correlation between $D/d$ and particle arrangement.

Two-dimensional (2D) configurations are also frequently encountered in various engineering applications and natural phenomena. Examples include the flow of products on conveyor belts and traffic jams. In this work we conduct 2D experiments and DEM simulations of clogging in a vertical pipe. Using DEM simulations, we study the influence of the aspect ratio $D/d$, the number of  initial particles $N$, and the coefficient of friction $\mu$ on the clogging probability $J$. Our results show that even with disordered particle arrangements, a non-monotonic dependence of $J$ persists in 2D vertical pipes. Moreover, we find that the reduction in $J$ is determined not only by geometric arrangements but also by the non-equilibrium state of the force and torque.

 \begin{figure}[t]
  	\begin{center}
  	\includegraphics[width=8.cm]{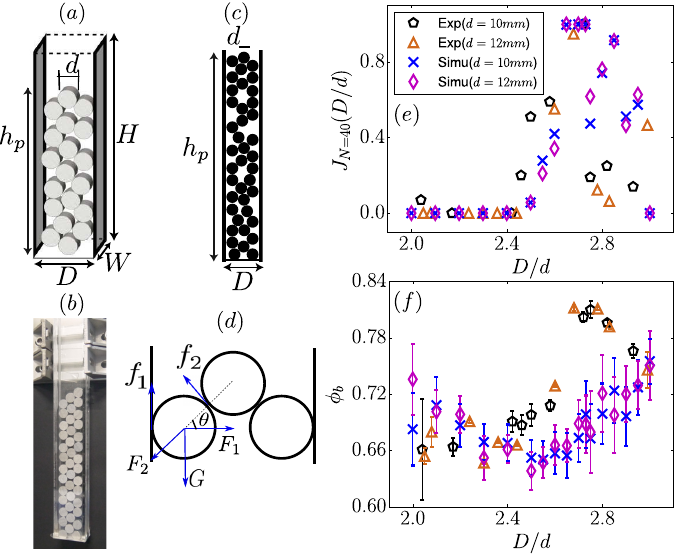}
	\vspace{-5pt}
	\caption{\label{Figure1:silo}(a) Sketch of experimental setup. (b) Photo of experimental setup. (c) Sketch of discrete simulation before the discharge process is initiated. (d) Diagram of a typical arch composed of three particles. Forces on one particle near the wall are represented with blue vectors, from left to right: the frictional force exerted by the wall $f_1$, normal force exerted by the central particle $F_2$ of weight $G$, normal force exerted by the wall $F_1$ and the frictional force exerted by the central particle $f_2$. Also shown is a comparison of the experimental and discrete simulation results: (e) clogging probability $J_N(D/d)$ and (f) initial bulk particle volume fraction $\phi_b$ versus $D/d$ for $N=40$ and $\mu=0.4$. The total number of trials for the data in (e) is reported in \cite{supplementary}. The results and error bars for $\phi_b$ in (f) are obtained from fewer trials ($N_t = 10$).}
	\vspace{-28pt}
  	\end{center}
\end{figure}

 \begin{figure*} 
  	\begin{center}
 \includegraphics[height=3.5cm]{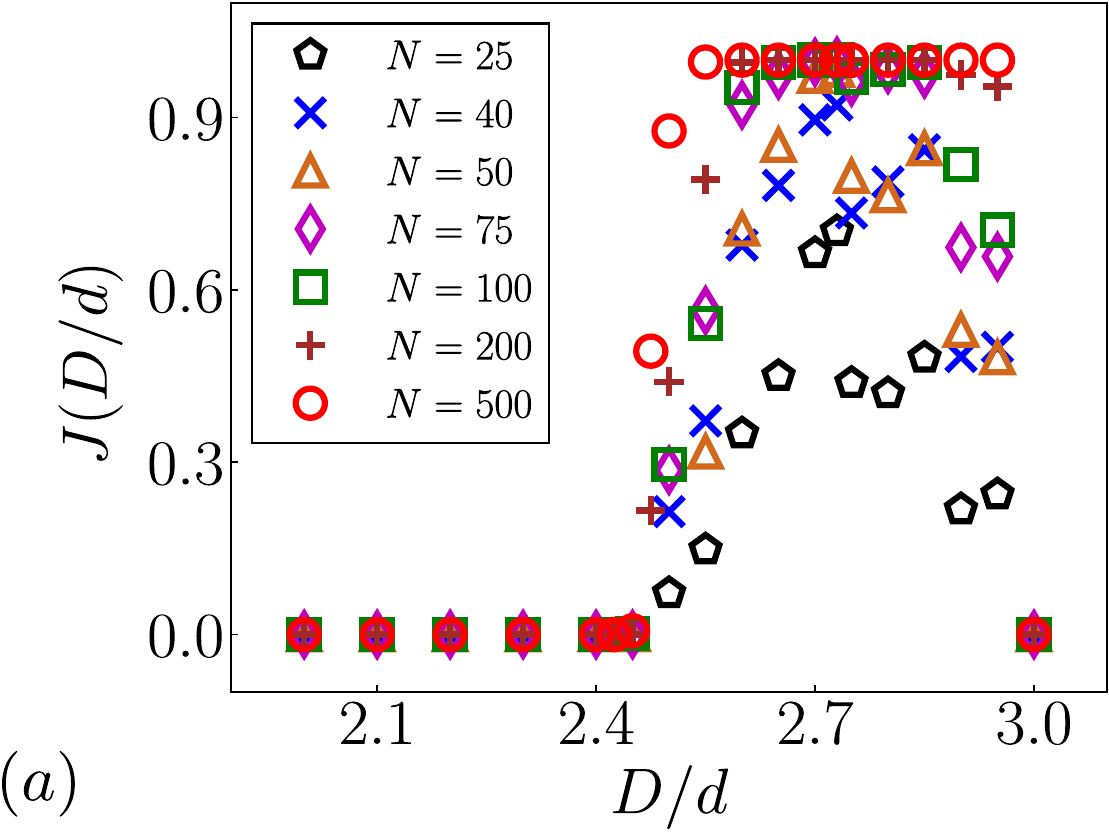}
\hspace{11pt} \includegraphics[height=3.5cm]{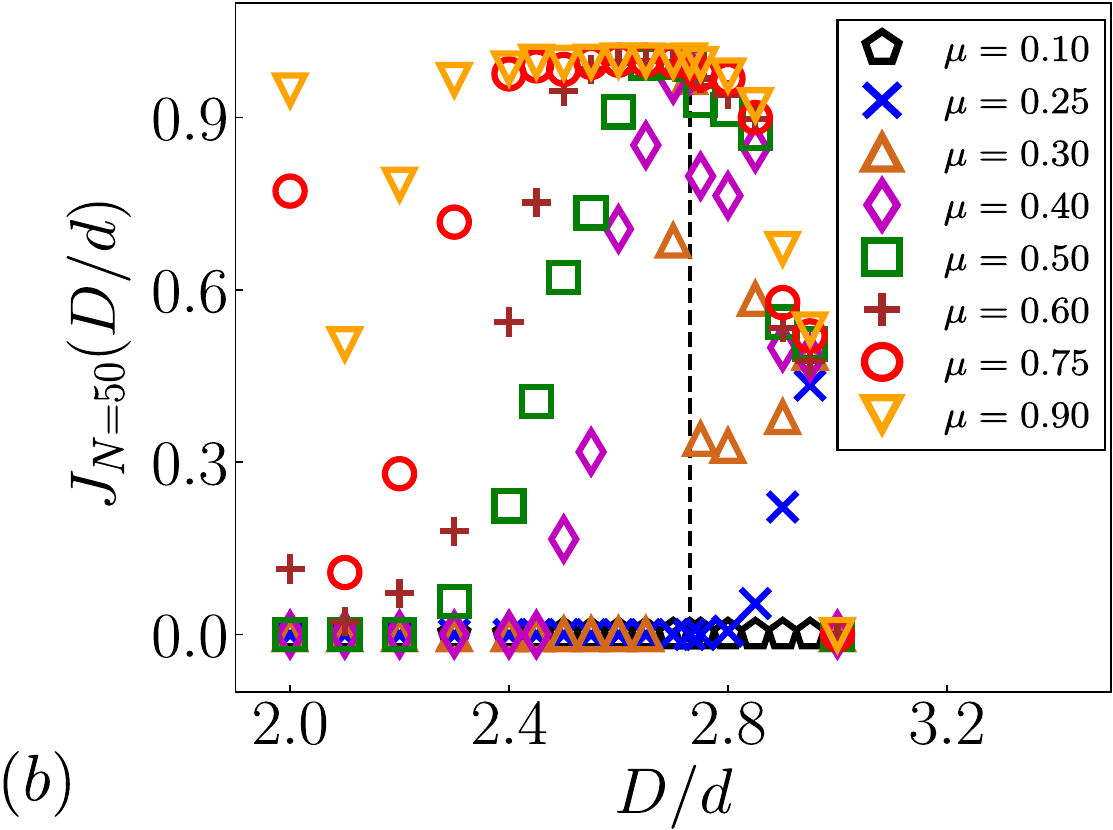}
\hspace{11pt}\includegraphics[height=3.55cm]{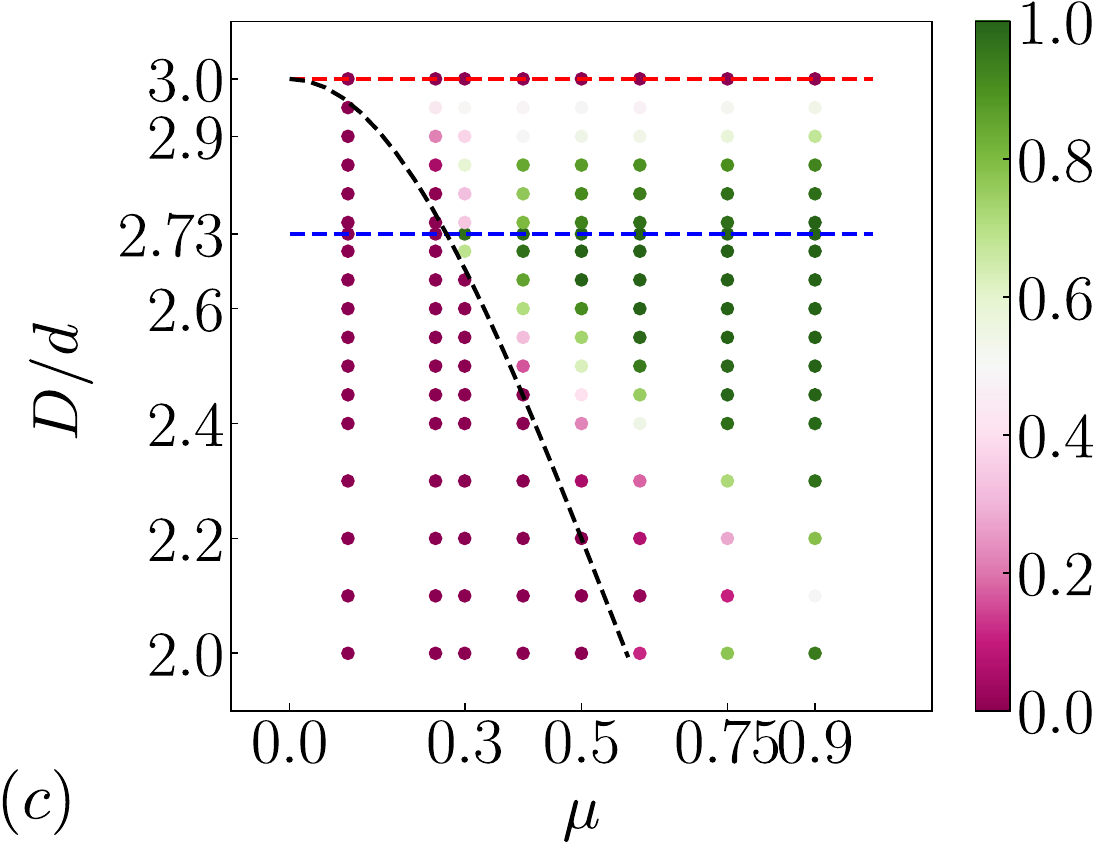}
    \vspace{-5pt}
  \caption{\label{Figure2:Jamprob}Discrete simulation results. Clogging probability $J_N(D/d)$ is plotted versus diameter aspect ratio $D/d$ (a) for various $N$ with $\mu=0.4$ and (b) for various $\mu$ with $N=50$. The vertical black dashed line represents $x=2.73$. (c) Clogging phase diagram of $D/d$ versus $\mu$. The black dashed line represents Eq.~\ref{equation8}, shown for $2 < y < 3$. Clogging probability follows the color scale.} 
  	\vspace{-15pt}
	\end{center}
\end{figure*}

\begin{figure*}[t]
   \begin{center}
     \includegraphics[height=4.cm]{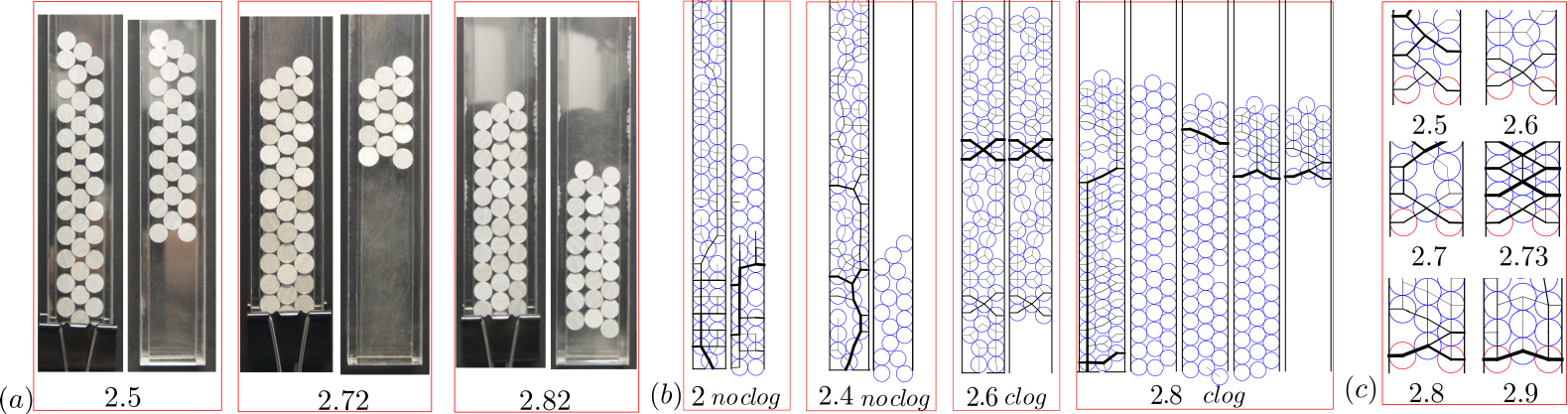}
      \vspace{-5pt}
	\caption{\label{Figure3:particlearrangement}(a) Typical particle arrangement at initial time and after clogging for various $D/d$ obtained by experiments. (b)Normal forces and particle arrangements for different $D/d$, obtained from simulations. Each panel shows the time sequence (from left to right) of the initial arrangement ($t = 0$) and subsequent states: for non-clogging cases at $t = 0.1875$ s ($D/d = 2.0, 2.4$); the clogged final state ($D/d = 2.6$); and the clogging evolution towards final arrest for $D/d = 2.8$, captured at $t = 0.025$, $0.055$, and $0.058$ s. The line thickness is proportional to the normal force. (c) Representative localized arch structure from a clogged state in the simulations.}
	\vspace{-25pt}			
  	\end{center}
\end{figure*}

\begin{figure*}[t]
   \begin{center}
       \includegraphics[height=3.7cm]{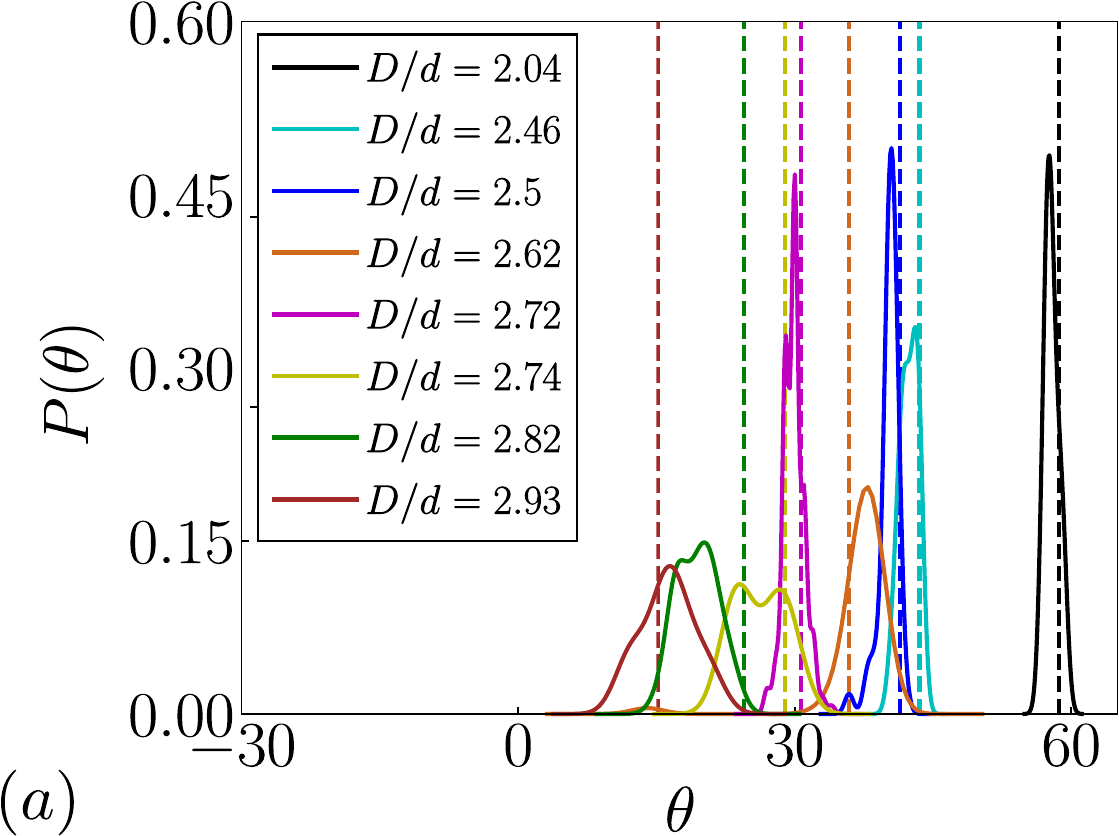}
 \hspace{11pt}\includegraphics[height=3.7cm]{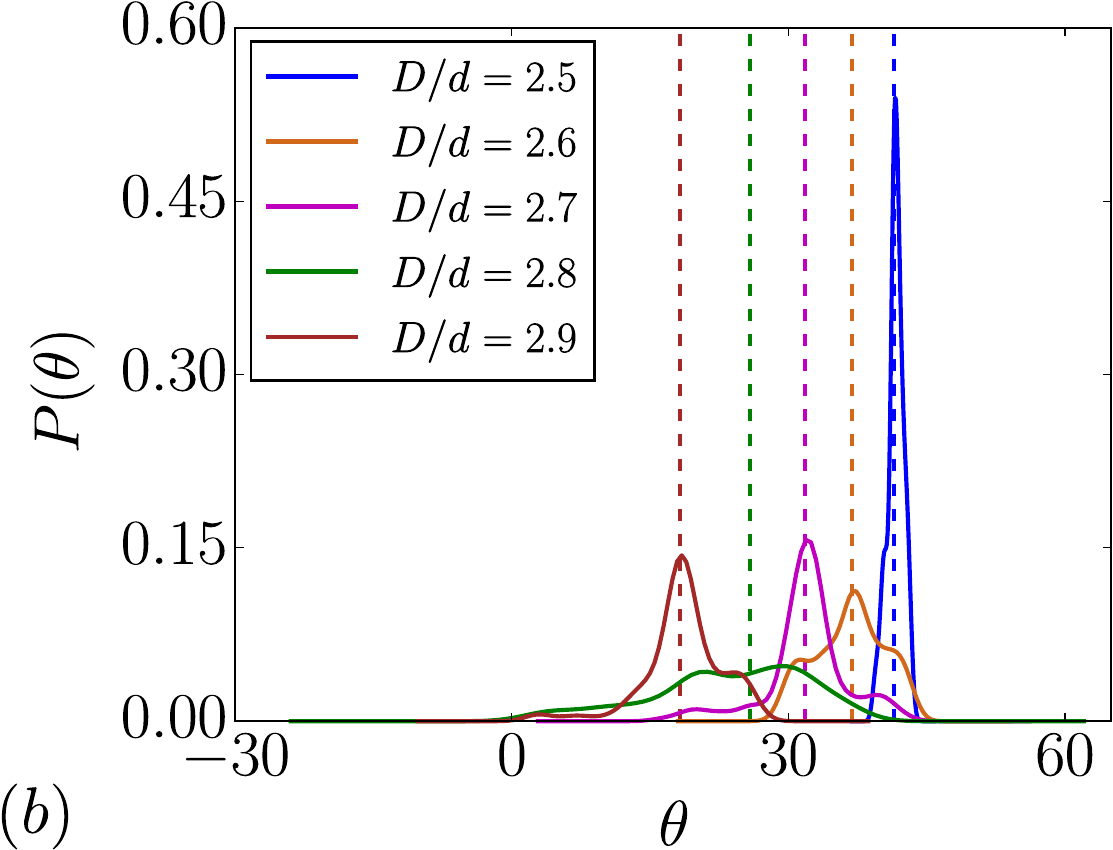}
  \hspace{9pt}  \includegraphics[height=3.6cm]{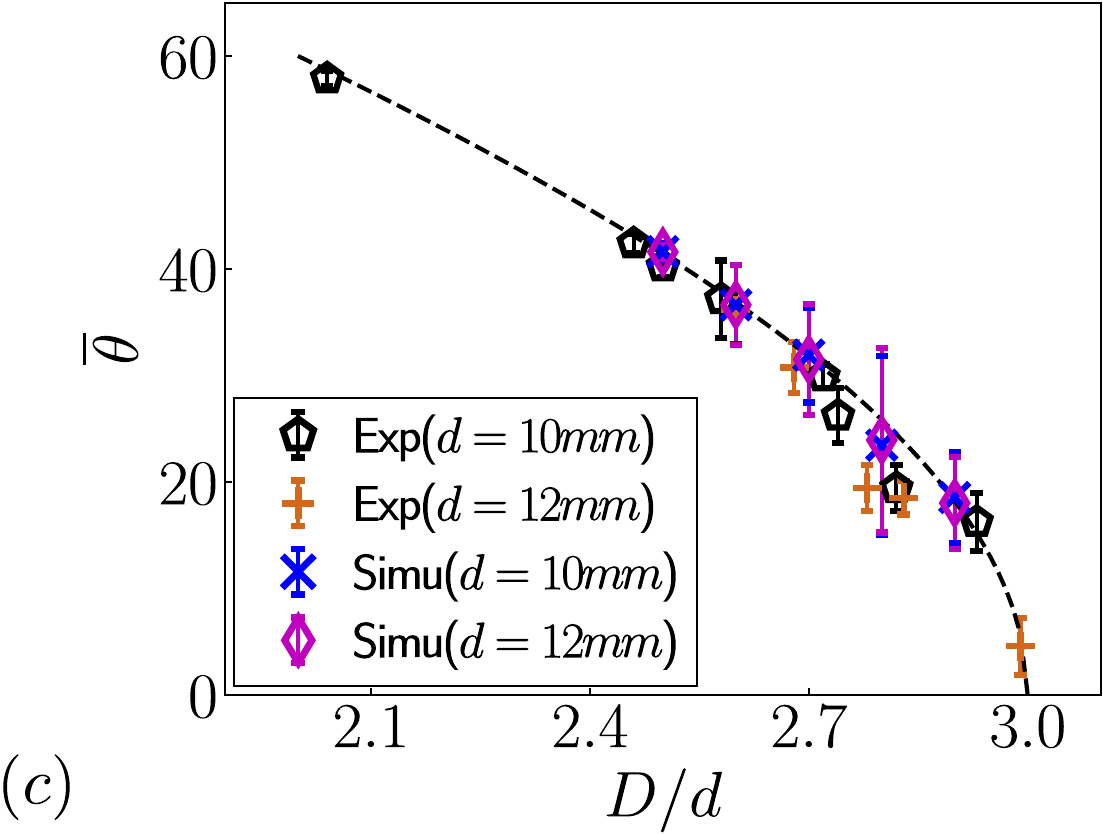}
    \vspace{-5pt}
	\caption{\label{Figure4:angle}Probability density distribution of the angle $\theta$ for various values of $D/d$: (a) experimental results with $d=10~mm$ and $N=40$ and (b) simulation results with $d=10~mm$, $N=40$ and $\mu=0.4$. (c) Average value of angle $\overline{\theta}$ versus $D/d$ for experiments and simulations. The dashed lines represent theoretical results of $\theta$ [Eq. \ref{equation1:theta}]. More than $300$ clogging events are generated in the simulations.}
	\vspace{-25pt}			
  	\end{center}
\end{figure*}

\textit{Methods.} We have set up an experimental device consisting of a 2D rectangular vertical pipe [see Figs. \ref{Figure1:silo}(a) and \ref{Figure1:silo}(b)]. The pipes are made of transparent polymethyl methacrylate. The typical height of the pipes $H$ is $300~mm$. The inner pipe diameters $D$ range from $20$ to $36~mm$.  The inner thickness of pipes $W$ is measured to be $7.5 \pm 0.1~mm$.  The granular sample consists of separate sets of monodisperse stainless steel disks, with diameters of $d=10 \pm 0.1~mm$ and $12 \pm  0.1~mm$, respectively. The thickness of the particles is smaller than the thickness of the pipe, which ensures a monolayer of particles in the rectangular vertical pipes. The experimental procedure begins by randomly and sequentially introducing a total number of disks $N=40$ from the top of the pipe with the aperture closed by a binder clip. Following this, the height of the particle bed, $h_p$, is measured to calculate the 2D volume fraction $\phi_b=N \pi d^2 /4 h_p D$. To initiate the flow, the binder clip is rapidly removed. In contrast to hopper flow, clogging in a vertical pipe can occur at any location, occasionally leading to multiple clogs. We employ the following definition: If all particles flow out of the pipe after the orifice is opened, the run is considered a nonclogging event; otherwise, if some disks are left in the pipe due to clogging, it is counted as a single clog event. If a clog occurs, we wait at least $30$ s to
confirm the persistence of the clog; in practice, transient clogs seldom occur and persist for a few seconds. For each set of parameters, we count the number of clogging events $N_c$ and obtain the clogging probability $J$ which is defined as $N_c/N_t$. Here $N_t$ is the number of the trials.  We adopt $N_t = 100$ and $300$ for disks with $d = 10$ and $12~mm$, respectively. For each clogging event, we capture an image of the arch closest to the bottom orifice with a standard HD camera Panasonic LUMIX. 

To further explore the underlying mechanism of the phenomenon, we also carry out DEM simulations to obtain more data and details. Following Zhou et al. \cite{Zhou2017} and Zou et al. \cite{Zou2020}, the LMGC90 software implementation of the contact dynamics method is used. The particles, interacting through a dense granular flow, are treated as perfectly rigid and inelastic. Contact dissipation is modeled in terms of a friction coefficient between particles $\mu_p$ and between particles and the walls $\mu_w$. Following Pournin et al. \cite{POU07}, we set $\mu_p=\mu_w=\mu$ to simplify the notation. 

To improve computational efficiency, the disks are initialized simultaneously with a prescribed inter-particle spacing. Then the granular column is prepared by randomly depositing particles in the closed vertical pipe, which minimizes the gravitational potential [see Fig. \ref{Figure1:silo}(c) for a typical initial arrangement]. Upon removal of the base plates, all particles begin to fall. In cases where all particles exit the pipe, the event is recorded as a non-clogging case, at which point the simulation is terminated; conversely, if the maximum particle velocity within the pipe drops below $10^{-6}$ m/s with remaining particles, the event is defined as a single clogging and the simulation ends. The $10^{-6}$ m/s criterion is chosen to eliminate the influence of transient clogs. The simulations use a time step of $\delta t = 5 \times 10^{-4}$ s and a maximum of $N_s = 2000$ steps. This duration is sufficient to ensure the complete passage of all particles in non-clogging scenarios. The $\phi_b$ and $J$ are determined using the same method as that employed in the experiment. 

The angle $\theta$ of the arch closest to the bottom orifice obtained with the experiment and simulation are both measured, as defined in Fig. \ref{Figure1:silo}(d). For both our experiments and simulations, all considered parameters are given in the Supplemental Material\cite{supplementary}. 

\textit{Results and discussion.} The experimental and simulation results are compared in Figs. \ref{Figure1:silo}(e) and \ref{Figure1:silo}(f). Figure \ref{Figure1:silo}(e) shows $J$ versus $D/d$ for $N=40$. The nonmonotonicity reported in the 3D cylindrical pipe \cite{Lopez2020} is also observed in 2D results. Similarly, clogging occurs in the regions where $2.45<D/d<3$ and $D/d \approx 2$. Figure \ref{Figure1:silo}(f) presents $\phi_b$ versus $D/d$ for $N=40$. We can observe that the experimental and simulation results of $J$ and $\phi_b$ are almost consistent for $D/d<2.73$, whereas a larger discrepancy arises when $D/d>2.73$. The differences between the experimental and simulation methods are discussed in \cite{supplementary}. Moreover, the dissimilar curve shapes of $J_N(D/d)$ and $\phi_b$ suggest that the correlation between the two is negligible. Subsequently, we can employ the DEM simulation approach to investigate the influence of different parameters on $J$.

Figure \ref{Figure2:Jamprob}(a) shows the dependence of $J$ on $D/d$ for different $N$.  As $N$ increases, $J$ also increases. We identify a critical value of the diameter aspect ratio ${(D/d)}_c \approx 2.45$ that separates clogging and nonclogging regimes in the range $2 < D/d < 3$. However, it can be seen that the increase in $N$ does not affect the value of ${(D/d)}_c$. This may be explained by the sharp transition in $J$ around ${(D/d)}_c$. Our findings are consistent with the results of L\'opez et al. \cite{Lopez2020}. In the range of $2 < D/d < 2.4$, their system with monodisperse particles exhibited near-infinite flow durations ($4-5$ h), even as $N$ tended to infinity. Therefore, we also refer to this nonclogging state as the flowing regime. Figure \ref{Figure2:Jamprob}(b) shows $J$ as a function of $D/d$ for various $\mu$ when $N = 50$. The clogging-flowing transition depends strongly on the friction. The value of ${(D/d)}_c$ decreases when $\mu$ increases. We thus construct a clogging-flowing $D/d-\mu$ phase diagram within the $2 < D/d < 3$ range [see Fig. \ref{Figure2:Jamprob}(c)], in which we can observe a clear clogging-flowing transition (see the black dashed line,  which corresponds to a theoretical model that will be presented subsequently).

To shed light on this anomalous dependence on $D/d$, we examine the typical particle arrangements corresponding to different values of $D/d$ for both experiment and simulation, shown in Figs. \ref{Figure3:particlearrangement}(a) and (b), respectively. In 2D nonclogging cases, we observe no significant signs of particle crystallization. Furthermore, we observe that the majority of clogging events occur right after initiating the flow by removing the binder clip at the bottom of the pipe (see more simulation results for the free-flowing time in \cite{supplementary}). A representative localized arch structure from a clogged state in the simulations is depicted in Fig. \ref{Figure3:particlearrangement}(c). We find that all clogging is mediated by three-particle arches across the parameter range studied. The angle $\theta$ depends strongly on $D/d$. In Figs. \ref{Figure4:angle}(a) and \ref{Figure4:angle}(b), we plot the probability density distribution (PDF) of $\theta$ for various $D/d$ obtained from experimental and simulation results, respectively. It can be observed that $D/d$ has an impact on the width of the PDF curves. Specifically, for smaller $D/d$, the distribution is narrower. The distribution is wider for $D/d>2.5$, especially for $D/d = 2.8$. If we suppose that the arch is symmetrical, based on the geometric restriction of nonpenetrating disks, which constitutes a fundamental constraint in the RRWM theory reported in \cite{TO01, Alborzi2023}, we can obtain the following equation of the angle $\theta$ for three-particle arches:
   \vspace{-4pt}
\begin{equation}
    \theta = arccos(0.5D/d - 0.5)
    \label{equation1:theta}
\end{equation}

Figure \ref{Figure4:angle}(c) shows the average value of $\theta$ versus $D/d$ for experiments and simulations. As shown by the dashed lines in Fig. \ref{Figure4:angle}, the values of $\overline{\theta}$ are consistent with those derived from Eq. \ref{equation1:theta}. We also find that $\overline{\theta}$ is almost independent of $\mu$ (see \cite{supplementary}). This equation can explain the decrease in $J$ around $D/d=2.9$. By analyzing the geometric properties of arches, we find that once the angle between the particles $\theta$ is greater than $arccos(0.5D/d - 0.5)$, the space left for the particles on the other side is too large to form a three-particle arch. For $D/d=3$, the probability of forming a three-particle arch approaches zero. 

In what follows, we perform an analysis of the forces and torques on one particle near the wall in a typical clogging arch [see Fig. \ref{Figure1:silo}(d)]. Our theoretical analysis is based on two fundamental assumptions: (i) The neighboring particle is above the particle near the wall and (ii) the particle near the wall is contacted by only one other particle. The force balance equations are given as follows: 
   \vspace{-4pt}
\begin{eqnarray} 
	&&F_2 sin \theta + G = f_1 + f_2cos \theta \label{equation2}\\
	&&F_2 cos\theta +f_2 sin \theta = F_1  \label{equation3}\\
	&&f_1 \le \mu F_1 \label{equation4} \\ 
	&& f_2 \le \mu F_2  \label{equation5}
\end{eqnarray}

According to the balance of torque, we have
   \vspace{-4pt}
 \begin{equation}
   f_1 = f_2
    \label{equation6}
\end{equation}

According to Eqs. \ref{equation2}-\ref{equation6}, we obtain 
   \vspace{-4pt}
\begin{equation}
    \mu \ge \displaystyle\frac{1-cos\theta}{sin\theta} 
    \label{equation7}
\end{equation}
Combining with Eq. \ref{equation1:theta}, which accounts for the geometry of a three-particle arch ($2 < D/d < 3$), we obtain the stability criterion:
   \vspace{-4pt}
\begin{equation}
\displaystyle\frac{D}{d} \ge \max \left\{ \displaystyle\frac{3-\mu^2}{1+\mu^2},  2  \right\}
    \label{equation8}
\end{equation}

\begin{figure}[t]
  	\begin{center}
	\includegraphics[height=3.4cm]{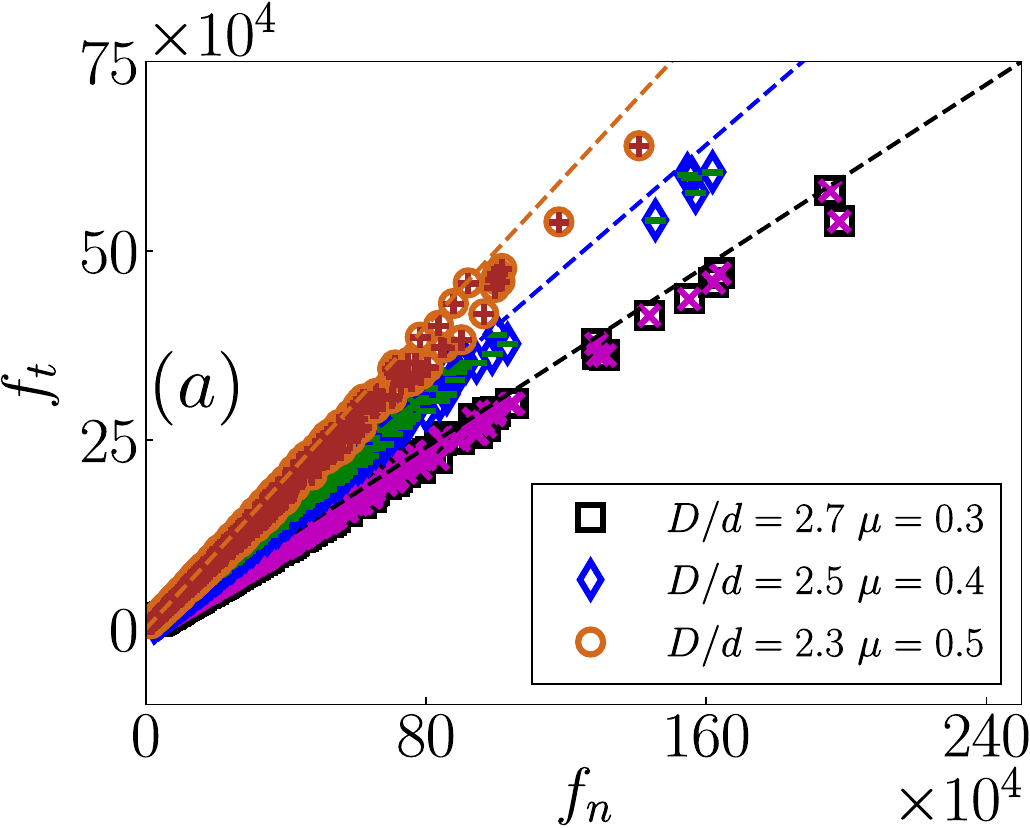}
	\includegraphics[height=3.2cm]{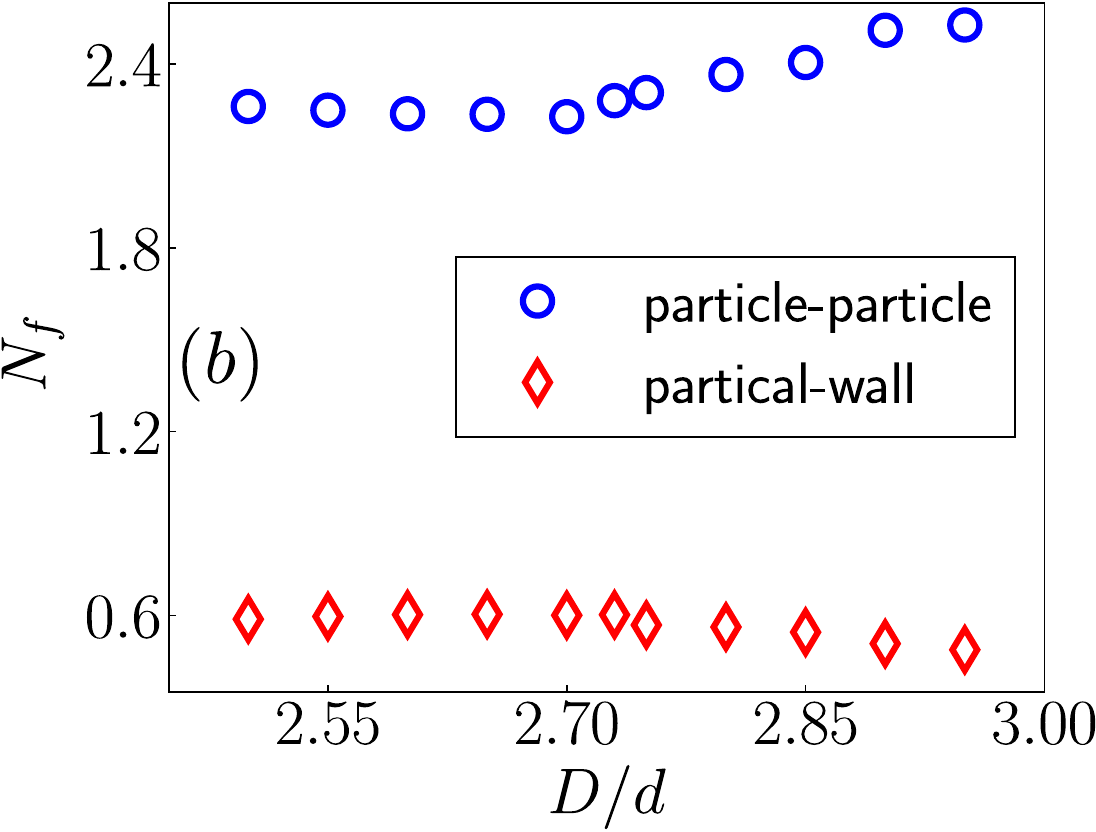}
	  \vspace{-5pt}
	\caption{\label{Figure5:force}(a) Tangential force $f_t$ versus normal force $f_n$ between particles in the arches for various $\mu$ and $D/d$. The dashed line represents the equation y = $\mu x$. (b) Number of contacts $N_f$ between particles and between particles and wall per particle in the bulk during the clogging event as a function of $D/d$ with $\mu=0.4$. More than $300$ clogging events are generated.}
       \vspace{-25pt}
  	\end{center}
\end{figure}

Then when $\mu=0.4$, we have $D/d \ge 2.45$. Interestingly, this is consistent with the critical clogging-flowing transition value $(D/d)_c$ obtained from experiments and simulations [see Figs.~\ref{Figure1:silo}(e) and \ref{Figure2:Jamprob}(a)]. The black dashed line in the clogging phase diagram [Fig. \ref{Figure2:Jamprob}(c)] represents Eq.~\ref{equation8}. We can see that below this line, there exists a flowing region, which validates this simple theoretical analysis. In contrast, above this line and when $D/d \gg (3-\mu^2)/(1+\mu^2)$, the system enters a geometrically controlled regime. Here, the primary determinant of clogging is the possibility of forming a three-particle arch, a condition that renders the effect of $\mu$ secondary [see Fig. \ref{Figure2:Jamprob}(b)]. 

We now discuss the scope and limitations of this theoretical approach. First, for the three-particle arch, the first assumption is always satisfied. Second, the force chain distribution reveals that the second assumption is primarily valid for $2 < D/d < 2.73$. At $D/d = 2$, particles can be subjected to significant horizontal forces from immediate neighbors, as illustrated in Fig.~\ref{Figure3:particlearrangement}(b) (see Fig. 3 in \cite{supplementary} for an example of a situation in which there is clogging). The horizontal forces enhance stability, making force balance more readily achieved than in configurations with $D/d > 2$, thereby explaining the second nonmonotonicity in $J$ observed at $D/d=2$ for $\mu \geq 0.6$ in Fig. \ref{Figure2:Jamprob}(b). Notably, we can see that the particles at the bottom near the wall are generally subjected to forces from one particle for $2<D/d<2.73$, while they become subjected to forces from two particles for $2.73<D/d<3$ [see red particles in Fig. \ref{Figure3:particlearrangement}(c)]. In this case, a greater tangential upward force is required to counterbalance the downward force and weight, making the clogging probability decrease abruptly at $D/d =2.73$ [see Fig. \ref{Figure2:Jamprob}(b)]. Furthermore, Fig. \ref{Figure4:angle} shows that as $D/d$ approaches $(D/d)_c$, the distribution of $\theta$ becomes remarkably narrow and converges to the value predicted by Eq. \ref{equation1:theta}, thereby validating the assumption of arch symmetry used in determining $(D/d)_c$. Therefore, our theoretical method is robust for predicting the flowing regime across $2 < D/d < 3$, although it slightly underestimates the actual range in the interval of $2.73 < D/d < 3$.

Figure \ref{Figure5:force}(a) shows the tangential force $f_t$ as a function of normal force $f_n$ between particles in the arches for the case in the clogging-flowing transition regime. Squares, diamonds, and circles represent the results of forces obtained by simulation, while pluses, dashes, and crosses represent the results of $f_t$ obtained by $f_t = (1-cos\theta) f_n /sin\theta $. The dashed line represents the equation $y=\mu x$. Near the clogging-flowing transition, we observe that $f_t$ becomes very nearly equal to but marginally less than $\mu f_n$. The results of $f_t$ obtained with simulation agree well with $f_t = (1-cos\theta) f_n /sin\theta $, demonstrating that the effective coefficient of friction is determined only by the actual angle. These results confirm that the occurrence of the flowing region in the clogging phase diagram can be attributed to the force and torque not being in equilibrium. In Fig. \ref{Figure5:force}(b) the numbers of contacts $N_f$ between particles and between the particles and wall per particle in the bulk during the clogging event are shown as a function of $D/d$. We can see that $N_f$ between particles (particles and wall) starts to increase (decrease) for $D/d \approx 2.73$. The increase in vertical forces and decrease in horizontal forces also explain the sudden drop in $J$ at $D/d = 2.73$.
 
\textit{Conclusion.} In summary,  the nonmonotonicity of $J$ found in cylindrical vertical pipe \cite{Lopez2020} is also observed in the 2D case. We have systematically proven that the occurrence of the flowing region for $2<D/d< (3-\mu^2)/(1+\mu^2)$ is attributed to the fact that the frictional stability criterion [Eq. \ref{equation8}] required for force and torque equilibrium is not achieved. A sudden decrease in $J$ occurs at $D/d=2.73$, which is due to the variations in the direction of contacts. A second significant decrease in $J$ occurs at around $D/d=2.9$, due to the geometric properties of arches as in silos \cite{TO01}.
 
In silos, the presence of the bottom wall ensures that all particles already satisfy the force and torque equilibrium even when $\mu = 0$. Consequently, the RRWM theory \cite{TO01}, which is based solely on the geometric characteristics of the arch, suffices to predict clogging probability in silos, as further evidenced in \cite{Garcimartin2010, Alborzi2023, Hathcock2025}. This explains why, in most cases, silos exhibit neither a non-monotonic dependence on $D/d$ nor a significant influence of the friction coefficient $\mu$ on the clogging probability \cite{POU07}.

In a future study we plan to extend this work to 3D and polydisperse systems over a wider $D/d$ range. It is noteworthy that this discontinuous clogging phenomenon has also been observed in tapered parallel microchannels \cite{Majekodunmi2022, Majekodunmi2023}. Future work could employ the statistical analysis methodology of Majekodunmi and Hashmi \cite{Majekodunmi2023} to better characterize the underlying dynamics.

\textit{Acknowledgments.} This work was sponsored by National Natural Science Foundation of China (Grant No. 12272134) and the Fundamental Research Funds for the Central Universities (Grant No. 2024MS052). Z. Z. thanks the National Natural Science Foundation of China for support (Grant No. 12302334). We acknowledge the assistance of DeepSeek in language polishing during manuscript preparation.

\bibliography{biblio}

\newpage

\clearpage
  
\onecolumngrid  
\begin{center}
   \bfseries\Large Supplementary material for "Clogging-flowing transition of granular media in a two-dimensional vertical pipe"
   \vspace{0.5em}
\end{center}
\twocolumngrid 

\setcounter{figure}{0}  
\setcounter{table}{0}  
\section*{PARAMETERS AND RESULTS DISCUSSION}
We provide the parameters used in experiments and simulations in Table \ref{table1} and Table \ref{table2}. Unless otherwise specified, the parameters reported correspond to all figures presented in this work.

Now we discuss the reasons for the differences between experimental and simulation results. Numerical and experimental trials are not conducted exactly in the same way. First, the experiment employs a rectangular vertical pipe in a quasi-2D configuration, whereas the DEM model utilizes a purely 2D domain. Second, the initial filling in the preparation phase is slightly different. The granular packing in simulations was generated by placing all particles simultaneously with a prescribed inter-particle gap before being released for gravitational free-fall inside the closed pipe. In contrast, the experimental packing was formed by adding particles randomly and sequentially from the top of the pipe with the aperture closed, resulting in a denser packing (see Fig. 1(e) in the main manuscript). Although mimicking the experimental filling in simulations would achieve comparable volume fraction $\phi_b$, it comes at a prohibitive computational cost. Nevertheless, we verified through the pair-correlation function that both initialization methods produce a disordered structure. Moreover, we have extensively verified, through both experiments and simulations, that the characteristic non-monotonic trend of clogging probability is reproducible with either of the two filling methods described above. Third, the numerical particle-wall friction coefficient is identical to the particle-particle friction coefficient, while friction occurs between plexiglas pipe and steel particles in the experiments.  However, using DEM simulations, we find that the clogging probability almost does not depend on the maximum value of particle-particle ($\mu_p$) and particle-wall ($\mu_w$) friction coefficients. Last, when the number of trials $N_t$ increases, particles are sliding (and rolling) on the pipe walls, which can cause erosions and importantly affect the friction coefficients, whereas this problem does not occur in simulations. According to probability theory, when $N_t$ exceeds $384$, the error margin of $J$ can be controlled within $5\%$. To balance the effects of erosion and errors arising from fluctuations in $J$ in experiments, while minimizing the influence of $N_t$, we adopt $N_t = 100$ and $N_t = 300$ for stainless steel disks with diameters $d = 10~mm$ and $d = 12~mm$, respectively.

In this work, we primarily focus on the clogging-flowing transition and nonmonotonicity of $J$ within the range $2 < D/d < 3$. Despite differences between experiment and simulation, our experimental results show that the clogging-flowing transition and nonmonotonicity are consistent with both our simulation and the experimental results obtained by L\'opez et al.\cite{Lopez2020}.  

\begin{table}
	\begin{center}
		\begin{tabular}{|c|c|c|c|}
			\hline
			$d~(mm)$ & Pipe inner diameter $D~(mm)$ & $N$ & $N_t$  \\
			\hline
			10	  &	20.4, 21.7, 24.2, 24.6, 2.5 &  40&	100\\
			 & 25.8, 27.2, 27.5, 28.2, 29.3 & &  \\
			 \hline
			12   & 24.6, 25, 26.9, 27.6, 28.3, 29.3, & 40 &	300\\ 
			   &  31.2, 32.2, 33.4, 34, 35.9 &   &	\\ 
			\hline
		\end{tabular}
	\end{center}
	\caption{\label{table1} Parameters used in experiments for given particle sizes ($d=10~mm$ and $d = 12~mm$), where $D$ is the pipe diameter, $N$ is the initial number of particles and $N_t$ is the number of trials.}
     \vspace{-5pt}
\end{table}

\begin{table}
	\begin{center}
		\begin{tabular}{|c|c|c|c|c|}
			\hline
			$d~(mm)$ & $D~(mm)$ & $\mu$ & $N$ & $N_t$\\
			\hline
			10	  &	[20, 21, 22,    & [0.1, 0.25, &  [25,40,  &	$\geq$ 500 \\
				  &	23, 24, 25,   & 0.3,0.4, &  50, 75, & when	\\        
				  &	25.5, 26, 26.5,  & 0.5,0.6, & 100,200, &  $N \leq 200$	 \\
				  &	 27, 27.3, & 0.75,0.9]& 500]  & 300	\\
				  & 27.5, 28, 28.5,  &&&when\\
				   & 29, 29.5, 30]  &&&$N=500$\\
			\hline
			12	  &	[24, 25.2, 26.4,  & 0.4 &  40  &	$\geq$ 500\\  
				  &	27.6, 28.8,30, &  &  &	\\
				  &	 30.6, 31.2, 31.8, & &   &	\\
				  &	32.76, 33, 34.2, & &   &	\\
				    &	33.6, 34.8,  & &   &	\\
				      & 35.4, 36] & &   &	\\
		        \hline
		\end{tabular}
	\end{center}
	\caption{\label{table2} Parameters used in DEM simulations for given particle sizes ($d=10~mm$ and $d = 12~mm$), where $D$ is the pipe diameter, $\mu$ is the friction coefficient, $N$ is the number of particles and $N_t$ is the number of trials.}
	 \vspace{-5pt}
\end{table}

 \begin{figure*}[ht]
  	\begin{center}
	(a)\includegraphics[height=4.cm]{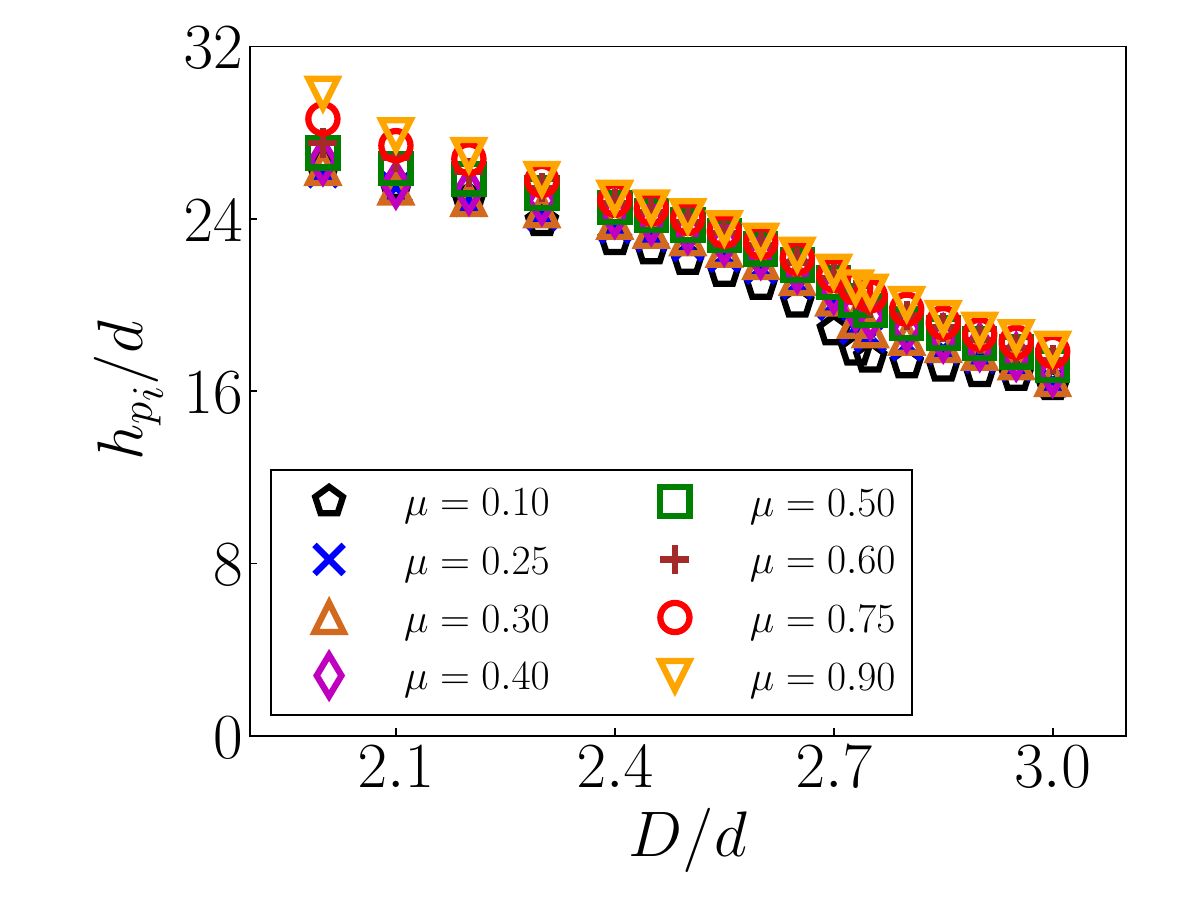}
 	(b)\includegraphics[height=4.cm]{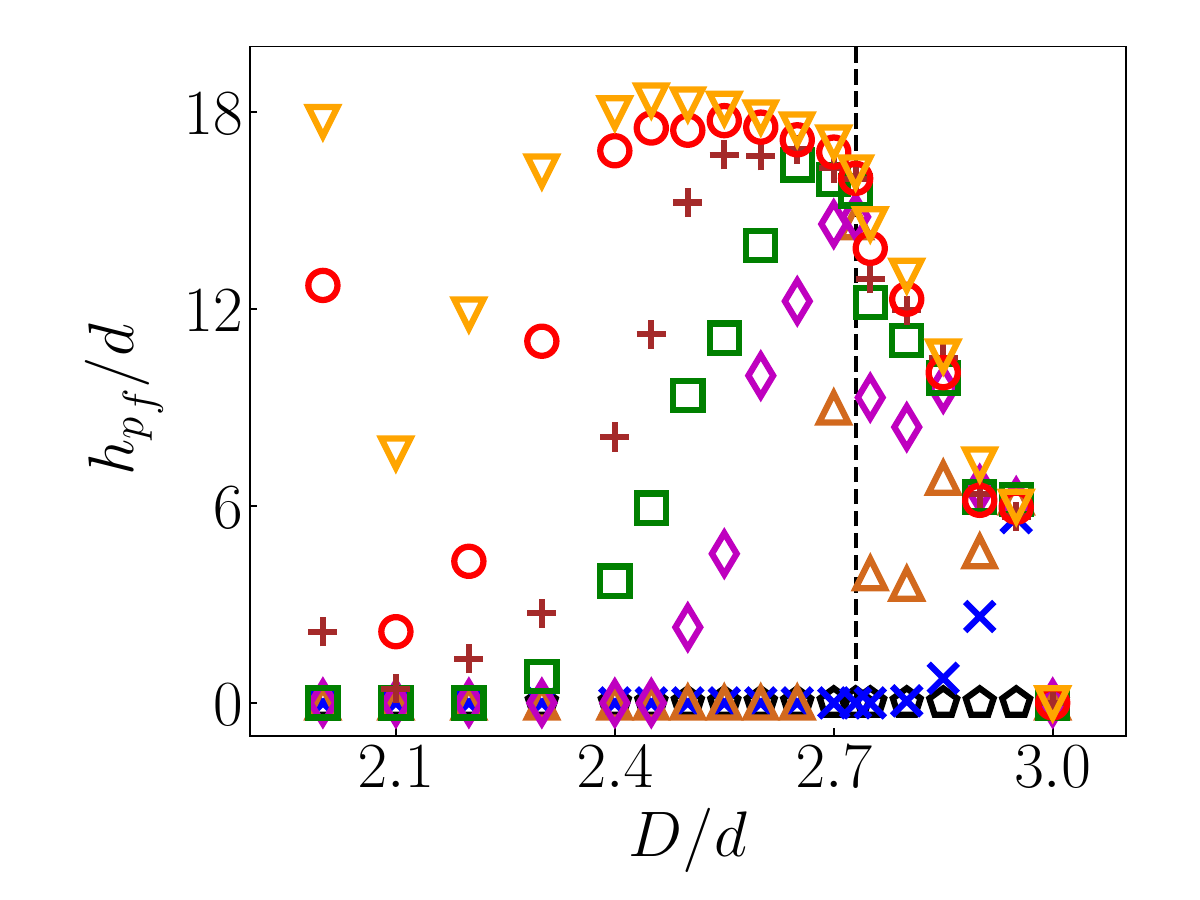}
  	(c)\includegraphics[height=4.cm]{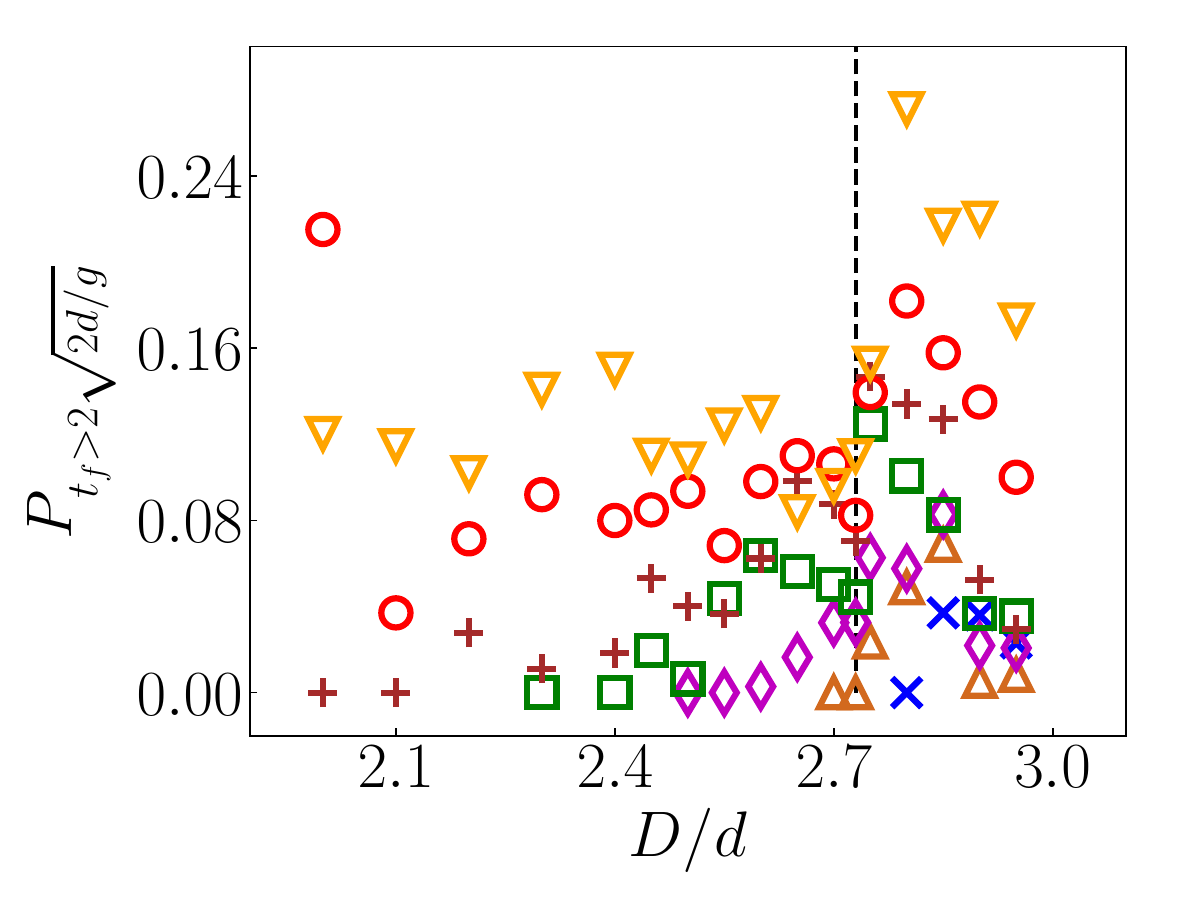}

	\caption{\label{Figures1} Discrete simulation results: (a) initial height of the particle bed ${h_p}_i$, (b) final height of the particle bed ${h_p}_f$, and (c) the free-flowing fraction of clogging events $P_{t_f>2\sqrt{2d/g}}$ versus diameter aspect ratio $D/d$ for various $\mu$ with $N = 50$. The dashed black lines in (b) and (c) represent $x=2.73$.
 }
 \vspace{-15pt}
  	\end{center}
\end{figure*}

 \begin{figure}[ht]
  	\begin{center}
	\includegraphics[height=5cm]{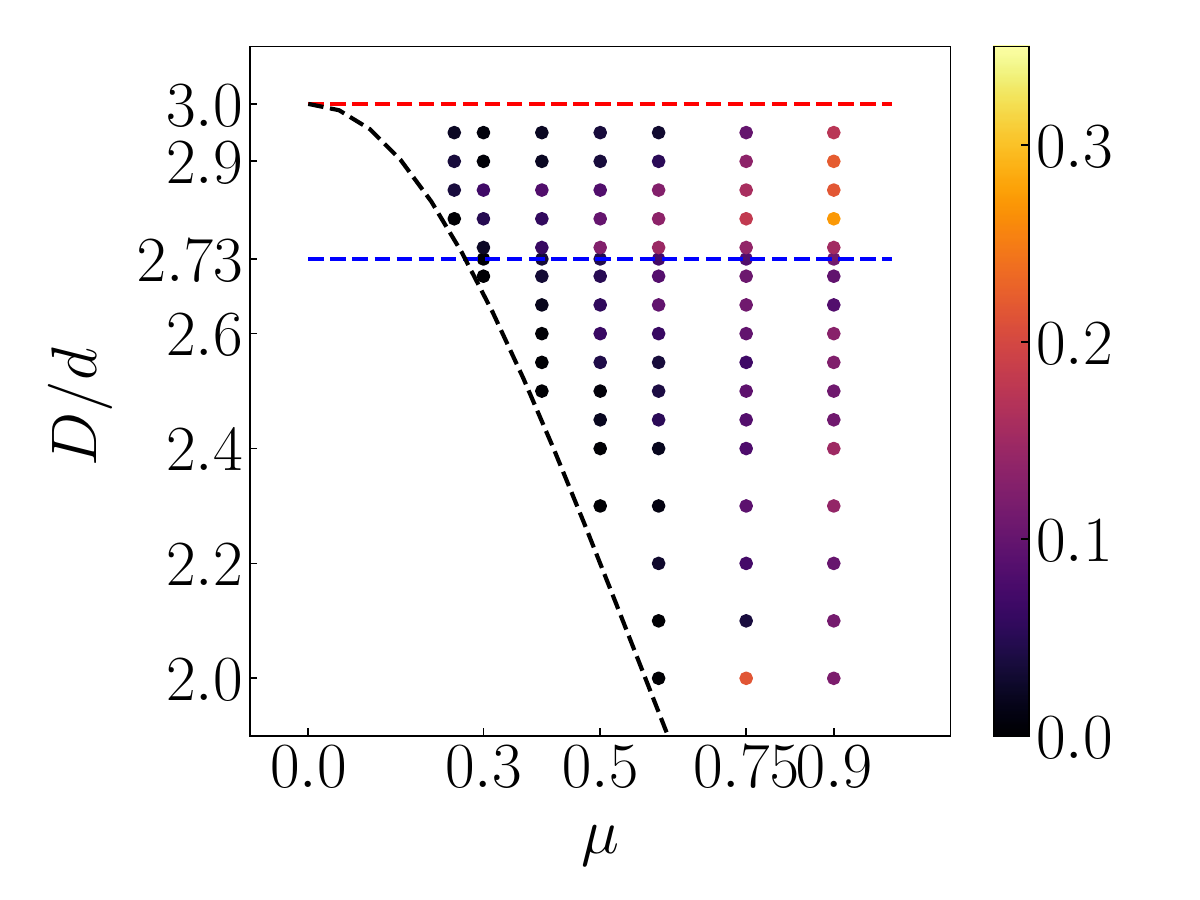}
	  \vspace{-5pt}
	\caption{\label{Figures2} A free-flowing fraction phase diagram $D/d-\mu$.  The dashed black line represents Eq.~8 in the main manuscript. The dashed blue and red horizontal lines represent $y = 2.73$ and $y = 3$, respectively. $P_{t_f>2\sqrt{2d/g}}$ follows the color scale.}
       \vspace{-15pt}
  	\end{center}
\end{figure}
 \begin{figure*}[ht]
  	\begin{center}
	 \includegraphics[height=6cm]{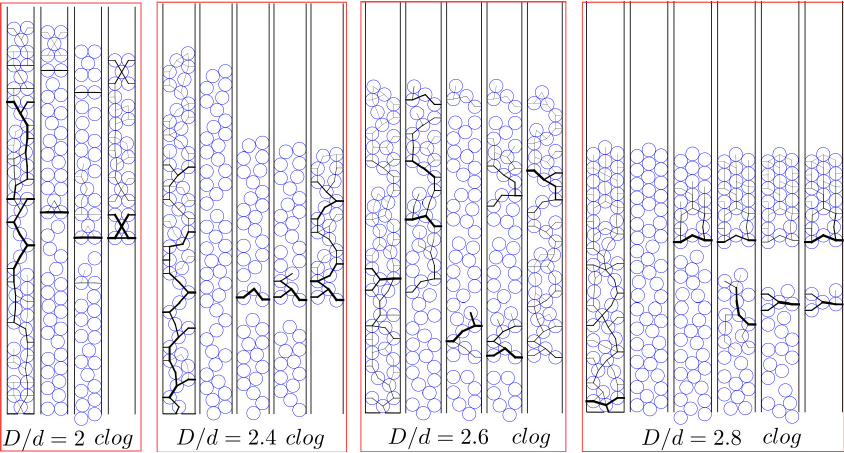}
	  \vspace{-5pt}
	\caption{\label{Figures3} Normal forces and typical particle arrangements for clogging case with free-flowing regime obtained with simulation results when $\mu = 0.75$. Each panel shows (from left to right) the initial arrangement $(t=0)$, the evolution process, and the final clogged state.  The evolution is captured at: $t = 0.050$ and $0.096$ s for $D/d = 2.0$; $t = 0.065$, $0.130$, and $0.134$ s for $D/d = 2.4$; $t = 0.032$, $0.064$, and $0.065$ s for $D/d = 2.6$; and $t = 0.025$, $0.050$, $0.075$, and $0.099$ s for $D/d = 2.8$. The line thickness is proportional to the normal force.}
       \vspace{-15pt}
  	\end{center}
\end{figure*}

\begin{figure}[!htbp]
  	\begin{center}
  	 \footnotesize{(a)}\includegraphics[height=5.cm]{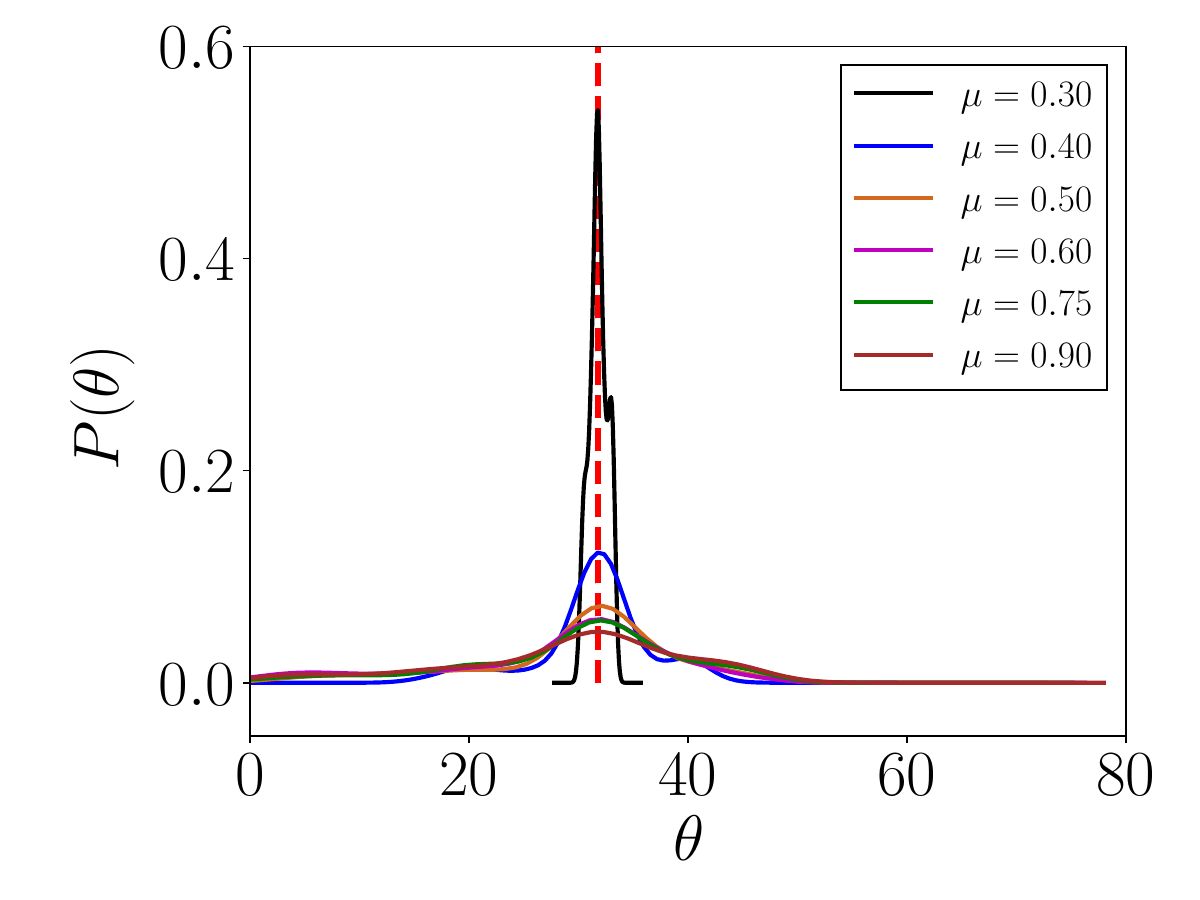}
	 \footnotesize{(b)}\includegraphics[height=5.cm]{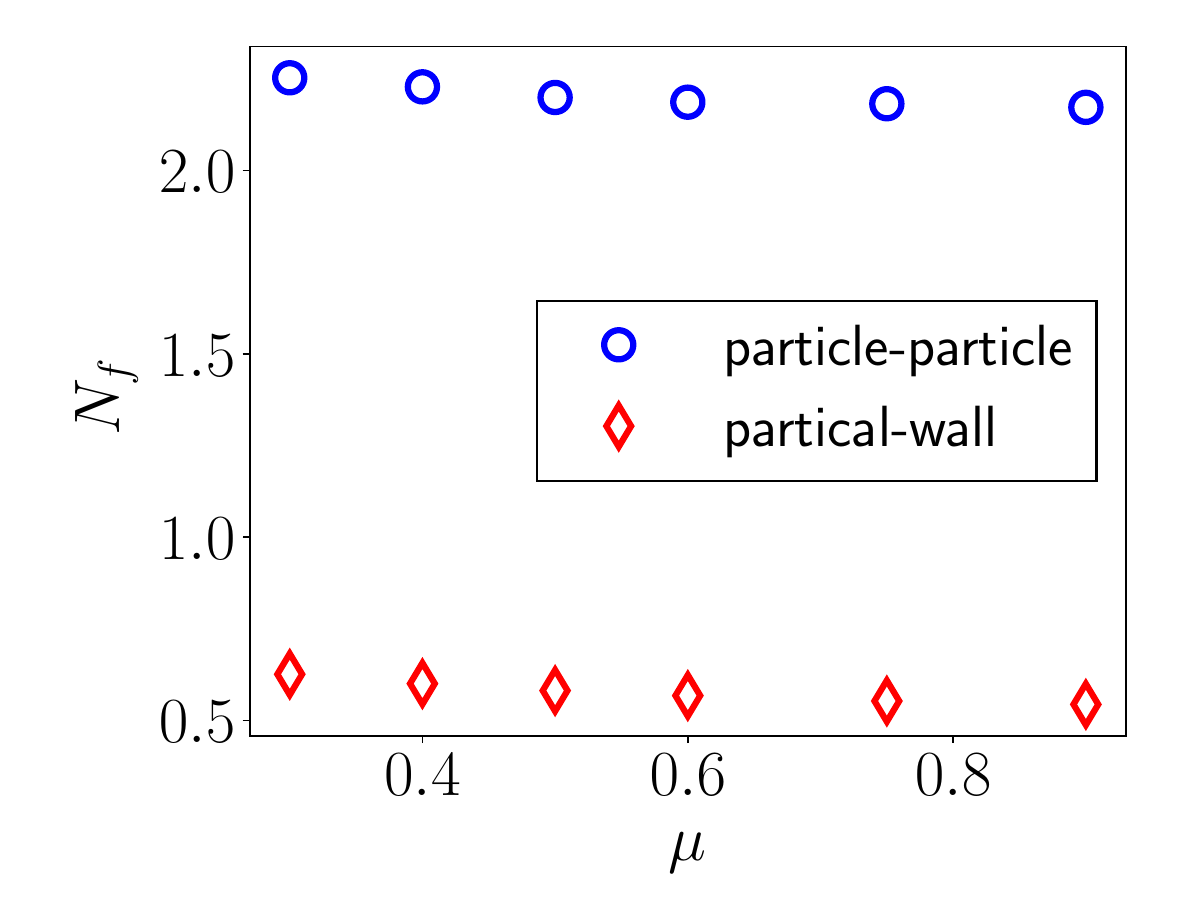}
	\caption{\label{Figures4} (a) Probability density distribution (PDF) of the angle $\theta$ for various $\mu$ when $D/d=2.7$. The dashed lines represent theoretical results of $\theta$ (Eq. 1 in the main manuscript). (b) The number of contacts $N_f$ between particle-particle and particle-wall per particle in the bulk during the clogging event as a function of friction coefficient $\mu$ when $D/d=2.7$. More than $300$ clogging events are generated.
}
 \vspace{-15pt}
  	\end{center}
\end{figure}
\section{STUDY OF  THE PARTICLE BED HEIGHT $h_p$ AND FREE-FLOWING DURATION $t_f$}

Besides the clogging probability $J$, parameters such as the height of the particle bed $h_p$ and the free-flowing duration $t_f$ are also examined in this work. The initial height of the particle bed ${h_p}_i$ and final height of the particle bed ${h_p}_f$ versus aspect ratio $D/d$ for $N=50$ obtained with DEM simulations are shown in Fig. \ref{Figures1}(a) and Fig. \ref{Figures1}(b), respectively. Fig. \ref{Figures1}(a) shows that ${h_p}_i$ decreases with $D/d$. Fig. \ref{Figures1}(b) and Fig. 2(b) in the main manuscript reveal that the final height of the particle bed ${h_p}_f$ and clogging probability $J$ share a similar trend. In contrast, the dissimilar curve shapes of ${h_p}_i$ and ${h_p}_f$ suggest that the influence of the particle bed ${h_p}_i$ is negligible compared to that of $\mu$ and $D/d$. Our experiments and simulations show that most clogging events occur near flow onset. The free-flowing duration $t_f$ is typically small when clogging occurs. This is due to the physical property that static friction is greater than kinetic friction. Therefore, we study the free-flowing fraction $P$ of clogging events (relative to the total clogging events $N_c$) that occur at the bottom of the particle bed, for which the free-flowing duration $t_f$ of the arch's leftmost particle is greater than $2\sqrt{2d/g}$ (representing a typical time it takes for the flow to traverse $2d$). The simulation results of $P_{t_f > 2\sqrt{2d/g}}$ versus $D/d$ for $N=50$ and various $\mu$ are shown in Fig. \ref{Figures1}(c). We can observe that as $D/d$ and $\mu$ increase, a free-flowing period can occur before clogging. Additionally, the free-flowing fraction $P_{t_f>2\sqrt{2d/g}}$ is observed to attain its maximum value near $D/d = 2.8$. Meanwhile, a second peak emerges near $D/d = 2$. We construct a free-flowing fraction phase diagram $D/d-\mu$ in Fig. \ref{Figures2}. $P_{t_f>2\sqrt{2d/g}}$ follows the color scale.  We can see that, near the clogging-flowing transition (dashed black line),  clogging occurs exclusively during the initial stage of flow. Because within this regime, force balance can only be achieved through static friction. When the system enters a geometrically controlled regime (i.e., $D/d \gg (3-\mu^2)/(1+\mu^2)$), force balance can also be attained through kinetic friction, the proportion of clogs that occur after free-flowing increases. 
\section{TYPICAL PARTICLE ARRANGEMENT DURING CLOG FORMATION WITH LARGE $\mu$} 
At $\mu = 0.4$, clogging is observed for $D/d > 2.45$, while free flow in clogging case does not appear until $D/d > 2.6$. Consequently, to better illustrate typical clog formation across the entire range of $D/d$ from $2$ to $3$, the typical particle arrangements and normal forces for clogging case with free-flowing regime obtained with simulation results when $\mu = 0.75$ are shown in Fig. \ref{Figures3}. 
We observe that clogging at $D/d=2$ is also initiated by the formation of a three-particle arch. In this configuration, the bottom particle near the wall experiences a significant horizontal force from a neighbouring particle. The horizontal force enables the particle near the wall to achieve force balance more readily. This explains the second nonmonotonicity in $J$ observed at $D/d=2$ for $\mu \geq 0.6$ (see Fig. 2(b) in the main manuscript). Moreover, a peak in the clogging probability at $D/d=2$ is also observed in the 3D experimental study by L\'opez et al.\cite{Lopez2020}. By comparing the cases in Fig. \ref{Figures3} and Fig. 3(b) in the main manuscript with different $\mu$ but the same $D/d$, we see that their particle configurations are very similar and both are in a disordered state. However, their clogging probabilities are significantly different. Clearly, the clogging probability of particles in 2D pipe case is determined not only by geometric arrangements but also by the non-equilibrium state of the force and torque.

\section{INFLUENCES OF $D/d$ AND $\mu$ ON OTHER PHYSICAL QUANTITIES} 

In this section, we extend our investigation to the influence of $D/d$ and $\mu$ on additional physical quantities: probability distribution function (PDF) of the angle $\theta$,  the number of contacts $N_f$ between particle-particle and particle-wall per particle in the bulk during the clogging event, the PDF of the mobilized friction acting on particles in the arches, and the normalized distribution of particle-particle normal forces.

\begin{figure}[t]
  	\begin{center}
     \footnotesize{(a)}   \includegraphics[height=5.cm]{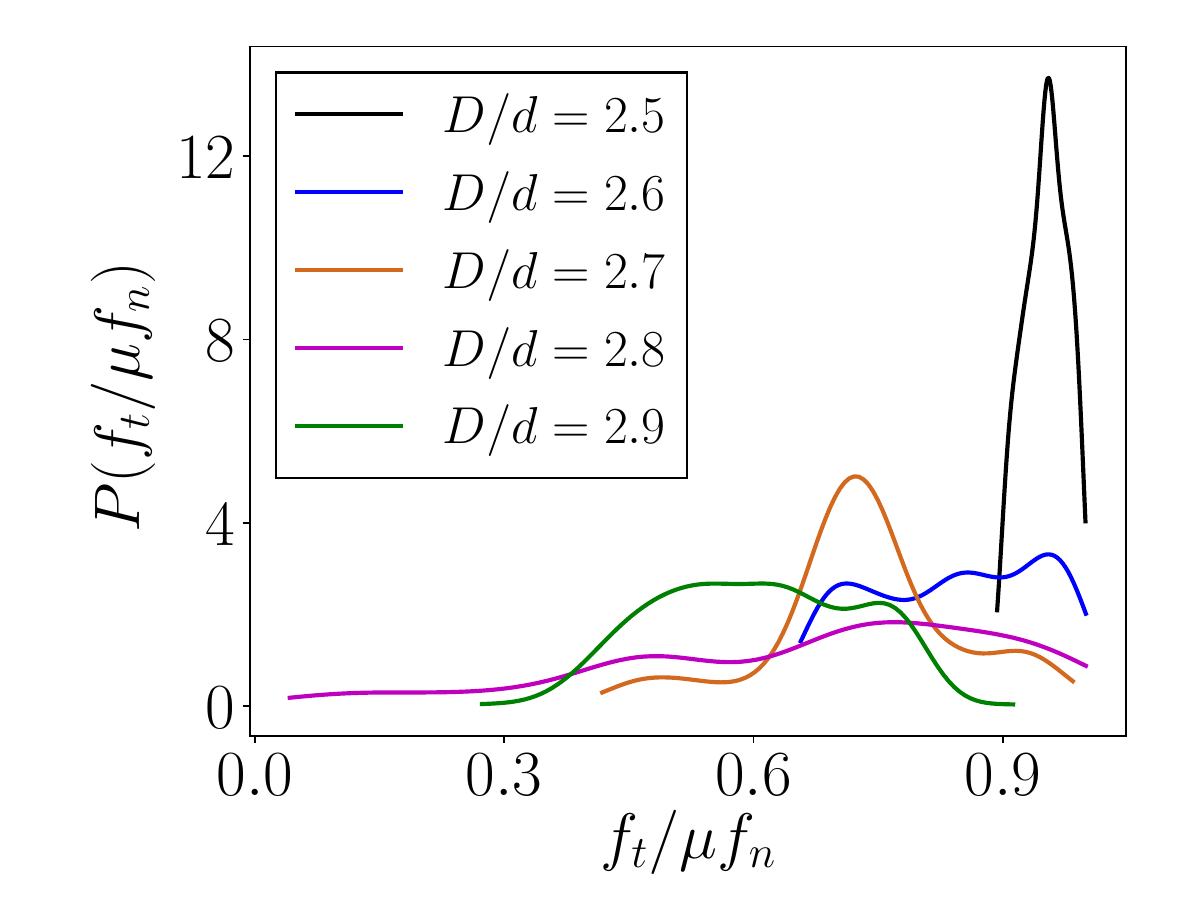}\\
     \footnotesize{(b)}   \includegraphics[height=5.cm]{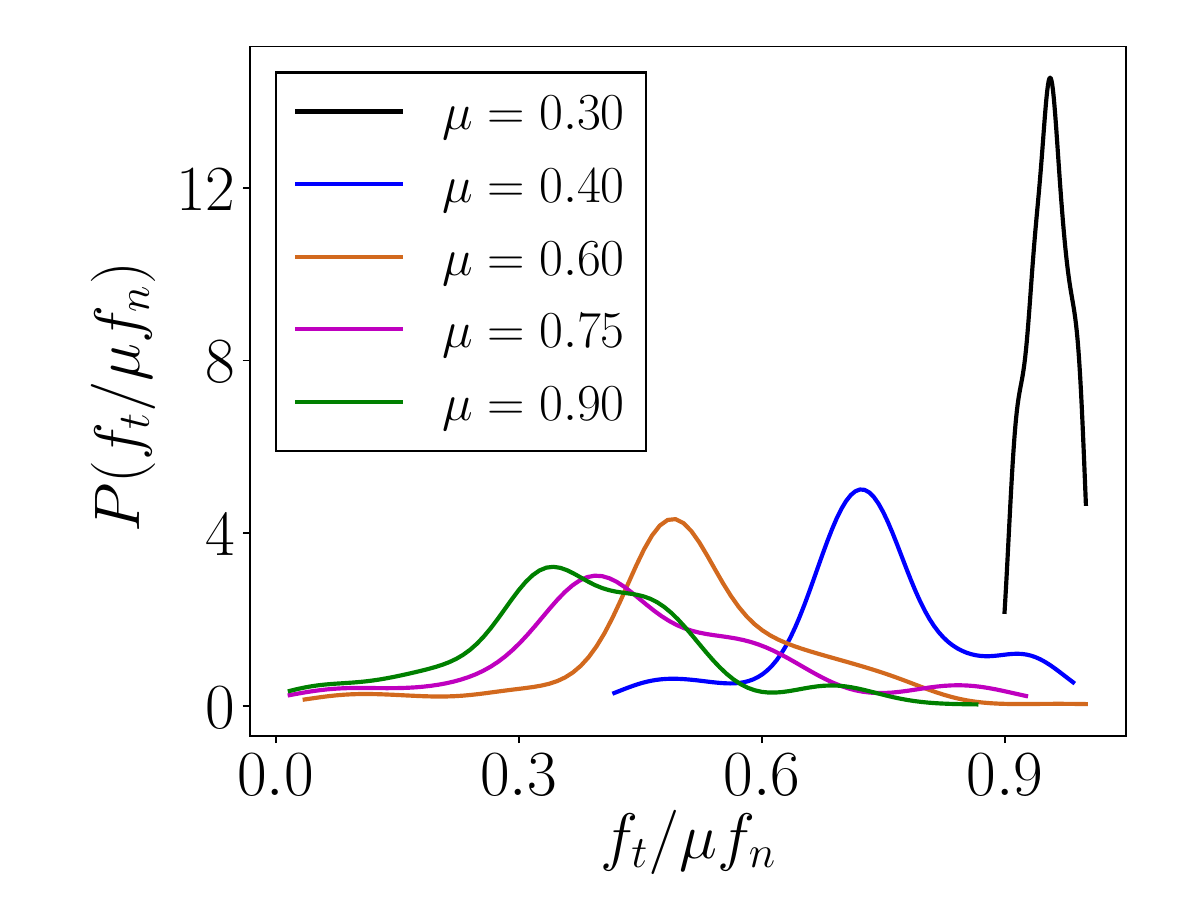}
				\caption{\label{Figures5} PDF of the mobilized friction  $f_t / (\mu f_n)$ of forces acting on particles at the arches for (a) various $D/d$ when $\mu=0.4$, (b) various $\mu$ when $D/d=2.7$. More than $300$ clogging events are generated.
				}
  	\end{center}
     \vspace{-15pt}
\end{figure}

\begin{figure}[t]
  	\begin{center}
	 \footnotesize{(a)}\includegraphics[height=5.cm]{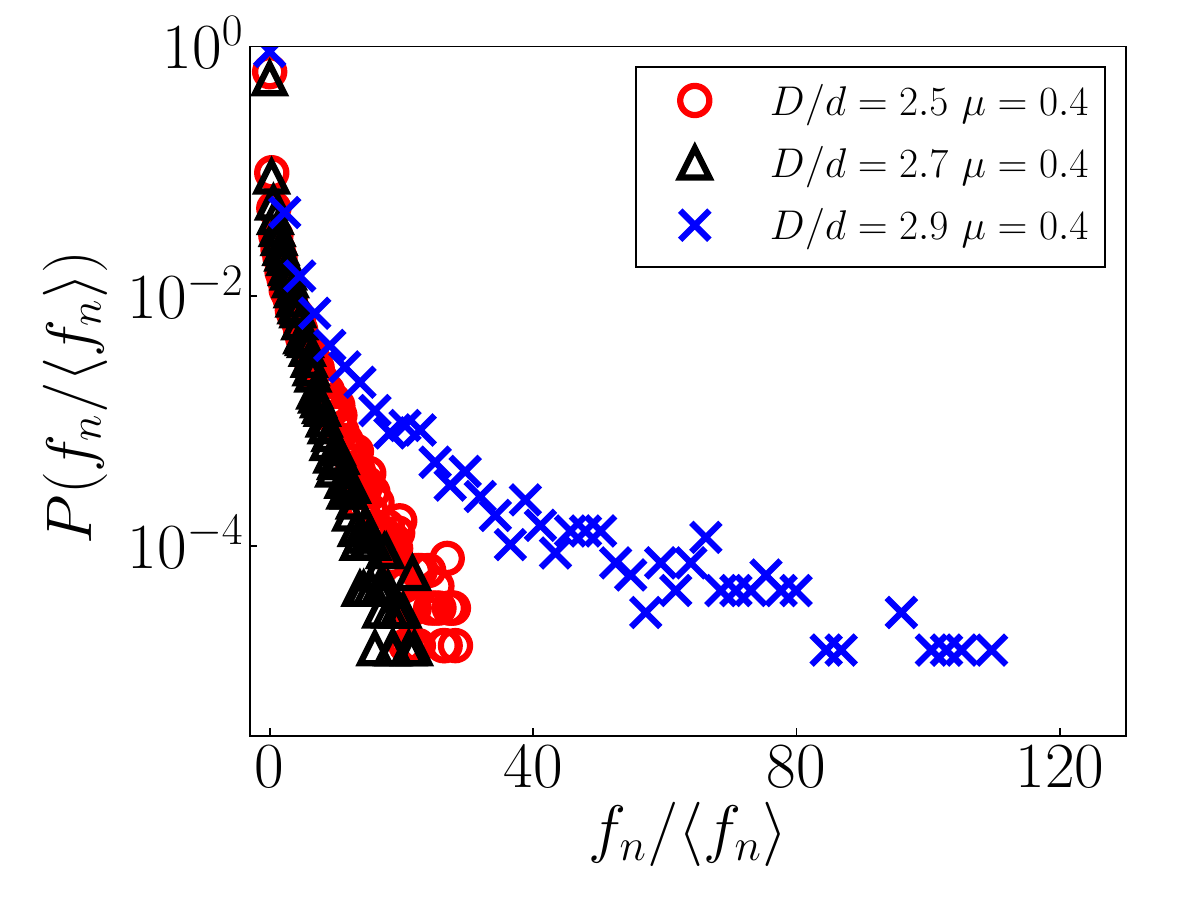}
	 \footnotesize{(b)}\includegraphics[height=5.cm]{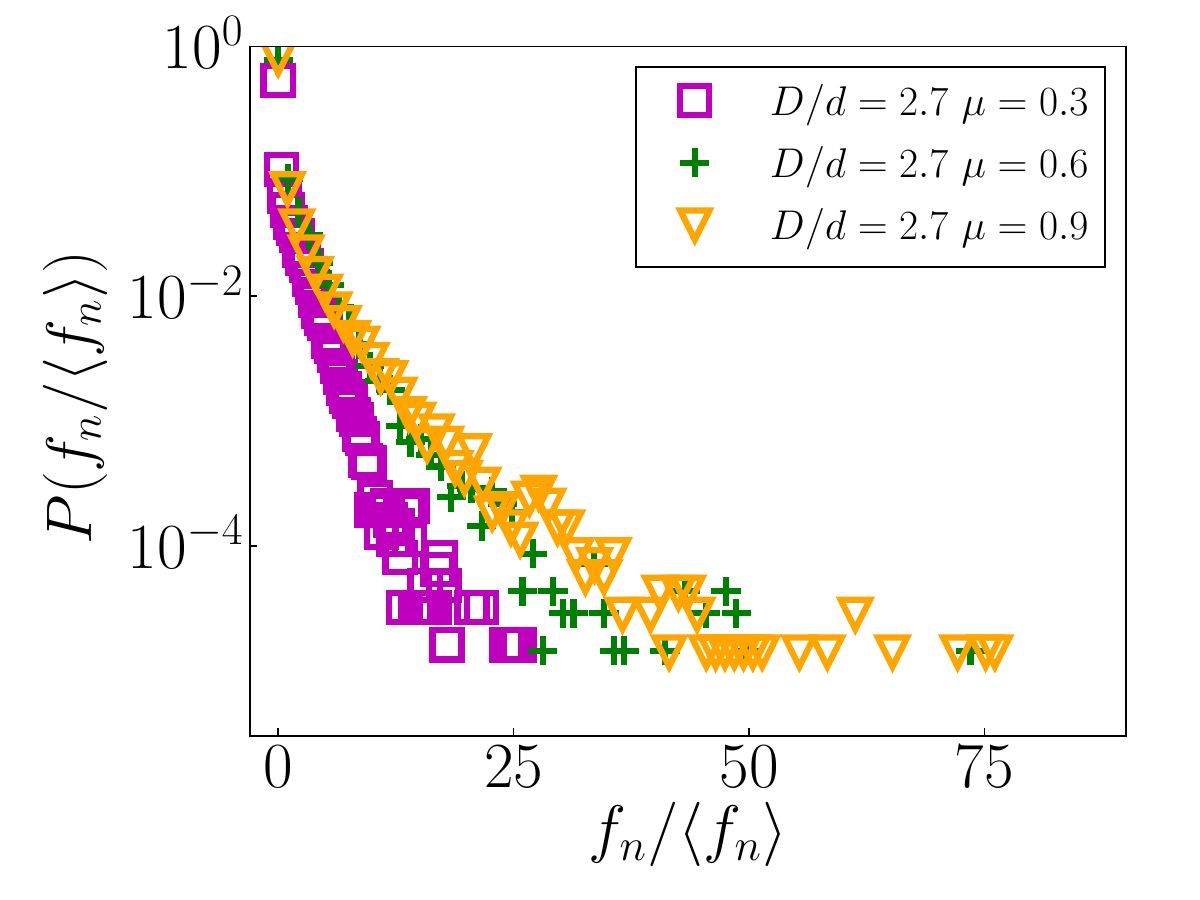}
	  \vspace{-5pt}
	\caption{\label{Figures6}  Normalized distribution of particle-particle normal forces $f_n/ \langle f_n \rangle$ in the bulk for (a) various diameter aspect ratio $D/d$ when $\mu=0.4$, (b) various $\mu$ when $D/d=2.7$. More than $1600$ clogging events are generated.
 }
       \vspace{-15pt}
  	\end{center}
\end{figure}

In Fig. \ref{Figures4}(a), we plot the PDF of $\theta$ for various values of $\mu$ when $D/d = 2.7$. It can be observed that as $\mu$ increases, the distribution broadens. The dashed lines in Fig. \ref{Figures4}(a) represent the theoretical value of angle obtained by Eq. 1 in the main manuscript, which further validates this equation. In Fig. \ref{Figures4}(b), the number of contacts $N_f$ between particle-particle and particle-wall per particle in the bulk during the clogging event is shown as a function of $\mu$ when $D/d = 2.7$.  It can be observed that $N_f$ exhibits a very slight decrease with increasing $\mu$, indicating a weak impact of $\mu$ on the force distribution.

Fig. \ref{Figures5} shows the distribution of the mobilized friction $f_t / (\mu f_n)$ of forces acting on particles at the arches for various $D/d$ and $\mu$. These distributions provide insight into how close the contacts should be to reach the Coulomb threshold criterion.  Fig. \ref{Figures5}(a) shows that as the ratio  $D/d$ increases, achieving the threshold condition $f_t =\mu f_n$ becomes more difficult.  Similarly, Fig. \ref{Figures5}(b) demonstrates that as the friction coefficient $\mu$ increases, it becomes more difficult to satisfy the threshold condition $f_t =\mu f_n$.  It is remarkable that when $\mu=0.3$ or $D/d=2.5$, the threshold condition $f_t =\mu f_n$  is nearly satisfied. This can be attributed to the fact that these cases are adjacent to the clogging-flowing transition in the clogging phase diagram (see dashed black line in Fig. 2(c) of the main manuscript and Fig. \ref{Figures2}).

Fig. \ref{Figures6} displays the normalized distribution of particle-particle normal forces in the bulk for various $D/d$ and $\mu$.  In 3D case, L\'opez et al.\cite{Lopez2020} found a notable broadening in the distributions of interparticle normal forces of the scenarios with higher clogging probability. In the 2D case, Fig. \ref{Figures6}(a-b) reveals that with increasing $D/d$ and $\mu$, the system enters a geometrically controlled regime (i.e., $D/d \gg (3-\mu^2)/(1+\mu^2)$), the particles exhibit broader force distributions. This is attributed to their greater diversity of geometrical configurations, as illustrated in Fig. 4(b) of the main manuscript and Fig. \ref{Figures4}(a). However, it should be noted that the width of the force tail does not exhibit a strictly positive correlation with clogging probability in 2D case (see Fig. \ref{Figures6}(a)). Although we have $J(2.7) > J(2.9)> J(2.5)$, it turns out that the force tail is narrowest when $D/d=2.7$, while being notably wide when $D/d = 2.9$. From Fig. 5(b) in the main manuscript, we can see that the number of contacts between particle-particle reaches its maximum (minimum) for $D/d = 2.9$ ($D/d = 2.7$). Since the forces between particle-particle are mainly in the vertical direction, and the total weight of the particles remains constant, the average value of particle-particle normal force $\langle f_n \rangle$ reaches its minimum (maximum) for $D/d = 2.9$ ($D/d = 2.7$). This could potentially be the reason why the tail of the normalized distributions of interparticle normal forces for $D/d=2.9$ ($D/d=2.7$) is the widest (narrowest). 

In summary, we can observe that as $D/d$ and $\mu$ increase, a free-flowing period can occur before clogging, the PDF curves of the angle $\theta$, the mobilized friction $f_t / (\mu f_n)$ and the normalized distribution of particle-particle normal forces become wider. We can conclude that when the system enters a geometrically-controlled regime (i.e., $D/d \gg (3-\mu^2)/(1+\mu^2)$), where force balance is achieved, the particles exhibit a greater diversity of geometrical configurations and broader force distributions. 

[1] D.  L\'opez, D. Hern\'andez-Delfin, R. C. Hidalgo, D. Maza, and I. Zuriguel, Clogging-jamming connection in narrow vertical pipes, Phys. Rev. E 102, 010902 (2020).

\end{document}